\documentclass[twocolumn,aps,superscriptaddress]{revtex4}
\usepackage{graphicx}
\usepackage{amssymb,amsfonts,amsmath}

\begin{document}
\title{Universal features in the photoemission spectroscopy of high temperature superconductors}
\author{J. Zhao}
\affiliation{Department of Physics, University of Illinois at Chicago, Chicago, IL 60607}
\affiliation{Materials Science Division, Argonne National Laboratory, Argonne, IL 60439}
\author{U. Chatterjee}
\affiliation{Materials Science Division, Argonne National Laboratory, Argonne, IL 60439}
\author{D. Ai}
\affiliation{Department of Physics, University of Illinois at Chicago, Chicago, IL 60607}
\author{D.G. Hinks}
\affiliation{Materials Science Division, Argonne National Laboratory, Argonne, IL 60439}
\author{H. Zheng}
\affiliation{Materials Science Division, Argonne National Laboratory, Argonne, IL 60439}
\author{G. Gu}
\affiliation{Physics Dept., Brookhaven National Laboratory, P.O. BOX 5000, Upton, NY 11973}
\author{S. Rosenkranz}
\affiliation{Materials Science Division, Argonne National Laboratory, Argonne, IL 60439}
\author {J.-P. Castellan}
\affiliation{Materials Science Division, Argonne National Laboratory, Argonne, IL 60439}
\author{H. Claus}
\affiliation{Materials Science Division, Argonne National Laboratory, Argonne, IL 60439}
\author{M. R. Norman}
\affiliation{Materials Science Division, Argonne National Laboratory, Argonne, IL 60439}
\author{M. Randeria}
\affiliation{Department of Physics, Ohio State University, Columbus, OH  43210}
\author{J. C. Campuzano}
\affiliation{Department of Physics, University of Illinois at Chicago, Chicago, IL 60607}
\affiliation{Materials Science Division, Argonne National Laboratory, Argonne, IL 60439}
\date{\today}

\begin{abstract}

The energy gap for electronic excitations is one of the most important characteristics of the superconducting state, as it directly reflects the pairing of electrons.  In the copper-oxide high temperature superconductors (HTSCs), a strongly anisotropic energy gap, which vanishes along high symmetry directions, is a clear manifestation of the $d$-wave symmetry of the pairing. There is, however, a dramatic change in the form of the gap anisotropy with reduced carrier concentration (underdoping). Although the vanishing of the gap along the diagonal to the square Cu-O bond directions is robust, the doping dependence of the large gap along the Cu-O directions suggests that its origin might be different from pairing. It is thus tempting to associate the 
large gap with a second order parameter distinct from superconductivity. We use angle-resolved photoemission spectroscopy (ARPES) to show that the two-gap behavior, and the destruction of well defined electronic excitations, are \emph{not} universal features of HTSCs, and depend sensitively on \emph{how} the underdoped materials are prepared. Depending on cation substitution, underdoped samples either show two-gap behavior or not. In contrast, many other characteristics of HTSCs, such as the dome-like dependence of $T_c$ on doping, long-lived excitations along the diagonals to the Cu-O bonds, energy gap at the antinode (crossing of the underlying Fermi surface and the ($\pi$, 0)-($\pi$, $\pi$) line) decreasing monotonically with doping, while persisting above $T_c$ (the pseudogap), are present in \emph{all} samples, irrespective of whether they exhibit two-gap behavior or not. Our results imply that universal aspects of high $T_c$ superconductivity are relatively insensitive to 
differences in the electronic states along the Cu-O bond directions.
\end{abstract}

\maketitle


Elucidating the mechanism of high temperature superconductivity in the copper oxide materials remains one of the most challenging open problems in physics. It has attracted the attention of scientists working in fields as diverse as materials science, condensed matter physics, cold atoms and string theory. To clearly define the problem of HTSCs, it is essential to establish which of the plethora of observed features are \emph{universal}, namely, \emph{qualitatively} unaffected by material-specific details. 

An important early result concerns the universality of the symmetry of the order parameter for superconductivity. The order parameter was found to change sign under a $90^{\circ}$ rotation \cite{DVH,TSUI}, which implies that the energy gap must vanish along the diagonal to the Cu-O bonds, i.e. the Brillouin zone diagonal. This sign change is consistent with early spectroscopic studies of near-optimally doped samples (those with the highest $T_c$ in a given family), where a $|\cos 2\phi|$ energy gap~\cite{SHEN,DING}  was observed ($\phi$ being the angle from the Cu-O bond direction), the simplest functional form consistent with $d$-wave pairing. More recently, there is considerable evidence~\cite{TANAKA,Hufner} that, with underdoping, the anisotropy of the energy gap deviates markedly from the simple $|\cos 2\phi|$ form. Although the gap node at $\phi = 45^\circ$ is observed at all dopings, the gap near the antinode (near $\phi = 0^\circ$ and $90^\circ$) is significantly larger than that expected from the simplest $d$-wave form.
Further, the large gap continues to persist in underdoped materials as the normal state pseudogap~\cite{Marshall, Nat96, Loeser} above $T_c$. This suggests that the small (near-nodal) and large (antinodal) gaps are of completely different origins, the former related to superconductivity and the latter to some other competing order parameter. 

This ``two-gap'' picture has  attracted  much attention \cite{Hufner}, raising the possibility that multiple energy scales are involved in the HTSC problem. There is mounting evidence for additional broken symmetries~\cite{Kaminski,Fauque,HeRH2011} in underdoped cuprates, once superconductivity is weakened upon approaching the Mott insulating state. The central issue is the role of these additional order parameters in impacting the universal properties of high Tc superconductivity. 

In this paper we use angle resolved photoemission (ARPES) to examine the universality of the ``two-gap'' scenario in HTSCs by addressing the following questions. To what extent are the observed deviations from a simple $d$-wave $|\cos 2\phi|$ energy gap independent of material details? How does the observed gap anisotropy correlate, as a function of doping, with other spectroscopic features such as the size of the antinodal gap, and the spectral weights of the nodal and antinodal quasiparticle excitations?

We systematically examine the electronic spectra of various families of cation-substituted Bi$_2$Sr$_2$CaCu$_2$O$_{8+\delta}$ single crystals as a function of carrier concentration
to elucidate which properties are universal, and which are not.
We present ARPES data on four families of float-zone grown Bi$_2$Sr$_2$CaCu$_2$O$_{8+\delta}$ single crystals, where $T_c$ was adjusted by both oxygen content and cation doping. As grown samples, labeled Bi2212, have an optimal $T_c$ of 91K. These crystals were underdoped to $T_c= 55$K by varying the oxygen content.  Ca-rich crystals (grown from material with a starting composition Bi$_{2.1}$Sr$_{1.4}$Ca$_{1.5}$Cu$_2$O$_{8+\delta}$) with an optimal $T_c$ of 82K are labeled Ca.  Two Dy-doped families were grown with starting compositions Bi$_{2.1}$Sr$_{1.9}$Ca$_{1-x}$Dy$_x$Cu$_2$O$_{8+\delta}$ with $x =$ 0.1 and 0.3 are labeled Dy1 and Dy2, respectively. A full list of the samples used and their $T_c$ determined from magnetization measurements are shown in the supplementary information, where we also show high resolution X-ray data that give evidence for the excellent structural quality of our samples.

Our main result is that the Dy1 and Dy2 samples show clear evidence of a two-gap behavior in the underdoped (UD) regime $T_c \lesssim 60$K, with loss of coherent quasiparticles in the antinodal region of ${\bf k}$-space where the gap deviates from a simple $d$-wave form. 
In marked contrast, the UD Bi2212 samples and the Ca samples show a simple $d$-wave gap in the superconducting state and sharp quasiparticles over the entire Fermi surface in a similar $T_c$ range of the underdoped regime. We conclude by discussing the implications of the non-universality of the two-gap behavior for the phenomenon of high $T_c$ superconductivity. 

We begin our comparison of the various families of samples by focusing in Fig.~1 on the superconducting state antinodal spectra as a function of underdoping. The antinode is the Fermi momentum \textbf{k}$_F$ on the $(\pi,0) - (\pi,\pi)$ Brillouin zone boundary, where the energy gap is a maximum and, as we shall see, the differences between the various samples are the most striking. We show data at optimal doping, corresponding to the highest $T_c$ in each family, in Fig.~1(a). 
Increasing Dy leads to a small suppression of the optimal $T_c$ compared to Bi2212, together with an increase in the
antinodal gap $\Delta_{max}$ and a significant reduction of the quasiparticle weight. This trend continues down to moderate underdoping, as seen in Fig.~1(b), where we show underdoped Bi2212 and Dy2 samples with very similar $T_c \simeq 66$K. 
For more severely UD samples, with $T_c \lesssim 60$K, spectral changes in the Dy-substituted samples are far more dramatic. In Fig.~1(c), we see that quasiparticle peaks in the Dy samples are no longer visible, even well below $T_c$, consistent with 
earlier work on Y-doped Bi2212 and also Bi2201 \& LSCO \cite{TANAKA,Terashima,Hashimoto,Kondo,Ma}. In contrast, Bi2212 and Ca-doped samples with comparable  $T_c$ continue to exhibit quasiparticle peaks. In this respect the latter two are similar to
epitaxially grown thin film samples that exhibit quasiparticle peaks all the way down to the lowest $T_c$ \cite{NODAL}.

A significant feature of the highly underdoped Dy samples in Fig.~1(c) is that, in addition to the strong suppression of the quasiparticle peak, there is severe loss of low energy spectral weight. To clearly highlight this, we show the doping evolution of antinodal spectra for the Dy1 (Fig.~1(d)) and Dy2 (Fig.~1(e)) samples. These observations are in striking contrast to the Bi2212 and Ca-doped data in Fig.~1(f), where we do see a systematic reduction of the quasiparticle peak with underdoping, but not a complete wipe out of the low-energy spectral weight. To the extent that the superconducting state peak-dip-hump line shape \cite{PEAKDIPHUMP, NORMANDING} originates from one broad normal-state spectral peak, the changes in spectra of the Dy materials are not simply due to a loss of coherence, but more likely a loss of the entire spectral weight near the chemical potential.

The doping evolution of the ${\bf k}$-dependent gap is illustrated in Figs.~2 and 3. In Fig.~2 we contrast the optimally doped Dy1 OP 86K sample (panels (a,b)) with a severely underdoped Dy1 UD 38K sample (panels (c,d)), the spectra being particle-hole symmetrized to better illustrate the gap. The OP 86K sample shows a well defined quasiparticle peak over the entire Fermi surface (panel (a)) with a simple $d$-wave gap of the form $\Delta_{\rm max}|\cos(2\phi)|$ (blue curve in panel (b)). For the UD 38K sample, we see in Fig.~2(c) well-defined quasiparticles near the node (red spectra), but not near the antinode (blue spectra). The near-nodal gaps (red triangles in panel (d)) are obtained from the energy of quasiparticle peaks and continue to follow a $d$-wave gap (blue curve in 2(d)). But once the quasiparticle peak is lost closer to the antinode, one has to use some other definition of the gap scale. We identify a ``break'' in the spectrum, by locating the energy scale at which it deviates from the black straight lines (panel (c)), which leads to the gap estimates (blue squares) in panel (d).

Despite the larger error bar associated with gap scale extraction in the absence of quasiparticles, it is nevertheless clear (Fig.~2(d)) that the UD 38K Dy1 sample has an energy gap that deviates markedly from the simple $d$-wave form. This observation is called ``two-gap'' in the UD regime, in contrast with a ``single gap'' near optimality (panel (b)). The two-gap feature found here is consistent with earlier work \cite{TANAKA,Terashima,Hashimoto,Kondo,Ma}. It is easy to observe from Fig.~2 that the Fermi surface angle at which the energy gap starts to deviate from $\Delta_{\rm max}|\cos(2\phi)|$ form matches with the one at which quasiparticel peak gets washed out. This is very similar to the two-gap behavior demonstrated in \cite{TANAKA,Terashima,Hashimoto,Kondo,Ma}. From these one might conclude that two-gap behavior is directly correlated with a loss of well-defined quasiparticle excitations in the antinodal region. However, we would like to point out the recently published ARPES data on Y doped Bi2212 \cite{Vishik_NJP, Vishik_PNAS}, where two-gap behavior has been observed despite the presence of sharp antinodal quasiparticle peaks.

We next show that the two-gap behavior is \emph{not} a universal feature of all underdoped samples. To make this point, we compare in Fig.~3 the gap anisotropies of the Ca-doped samples (panels (a,b)) with the Dy2 samples (panels (c,d)) with essentially identical $T_c$, where both families have the same optimal $T_c$. The near optimal samples, OD 79K Ca (panel (a)) and OP 81K Dy2 (panel (c)) both have a simple $d$-wave anisotropy (although different maximum gap values at the antinode).
But upon underdoping to similar $T_c$ values, the two have markedly different gap anisotropies. The UD 59K Dy2 sample (panel (d)) shows two-gap behavior, and an absence of quasiparticles near the antinode (similar to the discussion in connection with Fig.~2 above). However, the UD 54K Ca sample (panel (b)) continues to exhibit sharp quasiparticles and a ``single gap'', despite a very similar $T_c$ as the UD 59K Dy2. 

Having established the qualitative differences in the gap anisotropies for various samples as a function of underdoping, 
we next summarize in Fig.~4 the doping evolution of various spectroscopic features.  Instead of estimating the carrier concentration in our samples using an empirical formula \cite{Presland} (that may or may not be valid for various cation substitutions), we prefer to use the \emph{measured} $T_c/T_c^{\rm max}$ to label the doping. In Fig.~4(a) we show the doping evolution of the antinodal energy gap, which is consistent with the known increase in the gap with underdoping. 

The coherent spectral weight $Z$ for antinodal quasiparticles is plotted in Fig.~4(b) (for details on the procedure used to estimate this weight, from a ratio of spectral areas, see the supplementary information). The Dy1 and Dy2 samples both show a sudden and compete loss of $Z$ with underdoping \cite{Fournier}, which coincides with the appearance of two-gap behavior. In marked contrast to the Dy samples, the Bi2212 and Ca samples that exhibit a single $d$-wave gap show a gradual drop in the antinodal $Z$. On the other hand, we find that the nodal excitations are much less sensitive to how the sample is underdoped compared to the antinodal ones. Similar sharp nodal excitations have been observed in Dy doped Bi2212 samples in \cite{Vishik_PNAS} as well. The nodal quasiparticle weight $Z$ in Fig.~4(c) decreases smoothly with underdoping for \emph{all} families of samples,  as expected for a doped Mott insulator \cite{Anderson}. 

The two-gap behavior and the attendant loss of quasiparticle weight near the antinode imply a nodal-antinodal dichotomy, aspects of which have been recognized in ${\bf k}$-space \cite{Zhou,ShenKM,HeRH2009} and in real space \cite{McElroy,Kohsaka,Pushap}. Two possible, not mutually exclusive, causes of this behavior are disorder and competing orders. 

It is known that antinodal states are much more susceptible to impurity scattering, while near nodal excitations are protected \cite{Garg}. However, it is not a priori clear why certain cation substitutions (Dy) should lead to more electronic disorder than others (Ca). As shown by our X-ray studies in the supplement, there is no difference in the structural disorder in Dy and Ca samples. One possibility is that Dy has a local moment, but there is no direct experimental evidence for this.   

The two-gap behavior in UD materials, with a large antinodal gap that persists above $T_c$, is suggestive of an order parameter, distinct from $d$-wave superconductivity, that sets in at the pseudogap temperature $T^*$. There are several experiments \cite{Kaminski,Fauque,HeRH2011} that find evidence for a broken symmetry at $T^*$. However it is not understood how the observed small, and often subtle, order parameter(s) could lead to large antinodal gaps of $\simeq 50 - 80$meV, with a loss of spectral weight over a much larger energy range (Fig.~1(d,e)). 

Now we would like to discuss the presence or absence of competing order parameters and their pertinence, if they exist, in HTSC systems based on our ARPES and X-ray measurements. Firstly, in our ARPES measurements on any of our BISCO 2212 samples, we have not found any direct evidence for CDW or any other density wave ordering in terms of zone folding. Secondly, our X-ray diffraction measurements did not provide any signature for additional superstructures expected as a consequence of density wave ordering. However, none of these null results provide definitive evidence for the absence of a density wave ordering, particularly in a short range form or just on the surface. On the other hand, in the previously published works \cite{TANAKA,Terashima,Hashimoto,Kondo,Ma}, two-gap behavior has been conjectured to be a direct consequence of phase competition between $d$-wave superconductivity and some density wave ordering. As we have already demonstrated, two-gap behavior in underdoped samples is not a universal feature. Therefore, even if we assume that presence of competing order parameter is signified by two-gap behavior, it can't be central the superconductivity in HTSC systems.   

Whatever the mechanism leading to qualitatively different gap anisotropies for the UD Dy and Ca samples, it only produces relatively small, quantitative changes in key aspects of these materials, such as the dependence of $T_c$ on doping, the presence of sharp nodal quasiparticles, and the pseudogap. We thus conclude that antinodal states do not make a substantial contribution to the universal features of HTSCs. Clearly, two gaps are not necessary for high temperature superconductivity. Moreover, it is important to remember that samples with the highest $T_c$ always show a d-wave gap. The change in gap structure as the doping is reduced is not completely unexpected, as in strongly correlated samples, of which the HTSCs are prime examples, the relative strength of interactions depend sensitively on changes in doping and temperature, as shown by the many different phases exhibited in the phase diagram\cite{Vishik_PNAS, UTPALPNAS}.

\begin{acknowledgments}
We thank D. Robinson and J.P.C. Ruff for the support with the high energy X-ray diffraction measurements. Work at Argonne (J.Z., U.C., D.H., H.Z., S.R., H.C., M.R.N., and J.C.C.) was supported by UChicago Argonne, LLC, Operator of Argonne National Laboratory (ÒArgonneÓ). Argonne, a U.S. Department of Energy Office of Science laboratory, is operated under Contract No. DE-AC02-06CH11357. The U.S. Government retains for itself, and others acting on its behalf, a paid-up nonexclusive, irrevocable worldwide license in said article to reproduce, prepare derivative works, distribute copies to the public, and perform publicly and display publicly, by or on behalf of the Government.Basic Energy Sciences, . M.R. was supported by the DOE-BES  grant  DE-SC0005035. The Synchrotron Radiation Center is supported by the University of Wisconsin, Madison.
\end{acknowledgments}

\begin{figure}
\centerline{\includegraphics[width=.6\textwidth]{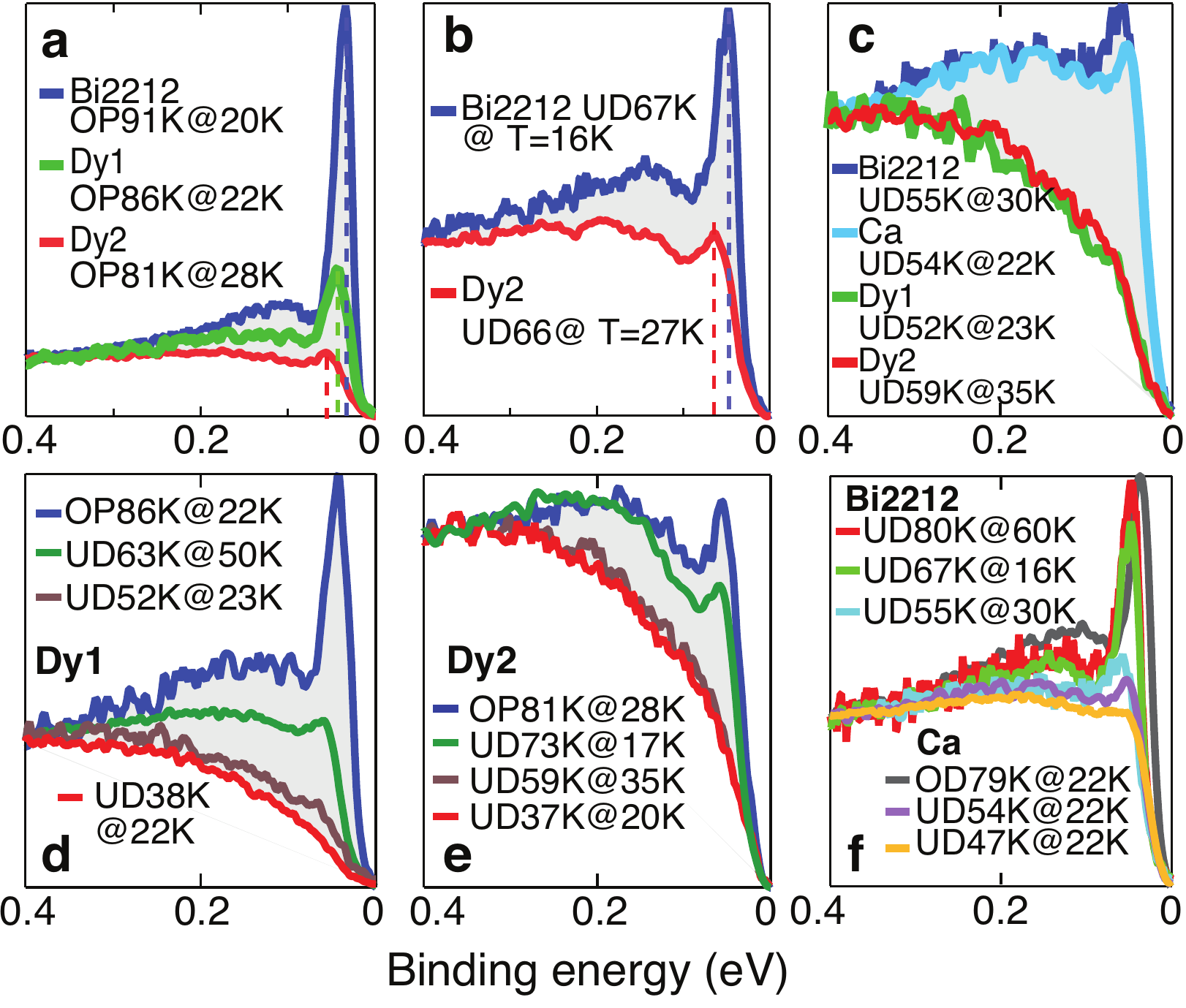}}
\caption{
Superconducting state antinodal ARPES spectra. We use the label ``Bi2212'' 
for samples without cation doping, ``Dy1'' for 10\%  Dy, ``Dy2'' for 30\% Dy,
and ``Ca'' for Ca-doped samples (see text). The temperature is indicated 
along with $T_c$. OP denotes optimal doped, UD underdoped and OD overdoped samples.
(a) Antinodal spectra for OP samples of three different families --
Bi2212 (blue), Dy1 (green) and Dy2 (red) -- showing an increase in gap and a decrease in 
quasiparticle weight with increasing Dy content.
(b) Antinodal spectra for UD samples with similar $T_c$ ($\simeq$ 66K) for
Bi2212 (blue) and Dy2 (red). As in (a), there is a larger gap and smaller
coherent weight in the Dy-substituted sample.
(c) Same as in (b), but for four UD samples with $T_c$ near 55K for
Bi2212 (dark blue), Ca (light blue), Dy1 (green) and Dy2 (red). The Bi2212 and Ca spectra
are very similar to each other and quite different from those of the Dy1 and Dy2 materials.
(d) Doping evolution of the antinodal spectra of four Dy1 samples from OP $T_c = 86$K to
UD $T_c = 38$K.
(e) Doping evolution of the antinodal spectra of four Dy2 samples from OP $T_c = 81$K to
UD $T_c = 37$K. We see in panels (d) and (e) the sudden loss of quasiparticle 
weight for $T_c$ below 60K.
(f)  Doping evolution of the antinodal spectra of three Bi2212 samples and three Ca samples, showing well-defined quasiparticle peaks in all cases.
}\label{fig1}
\end{figure}

\begin{figure}
\centerline{\includegraphics[width=.5\textwidth]{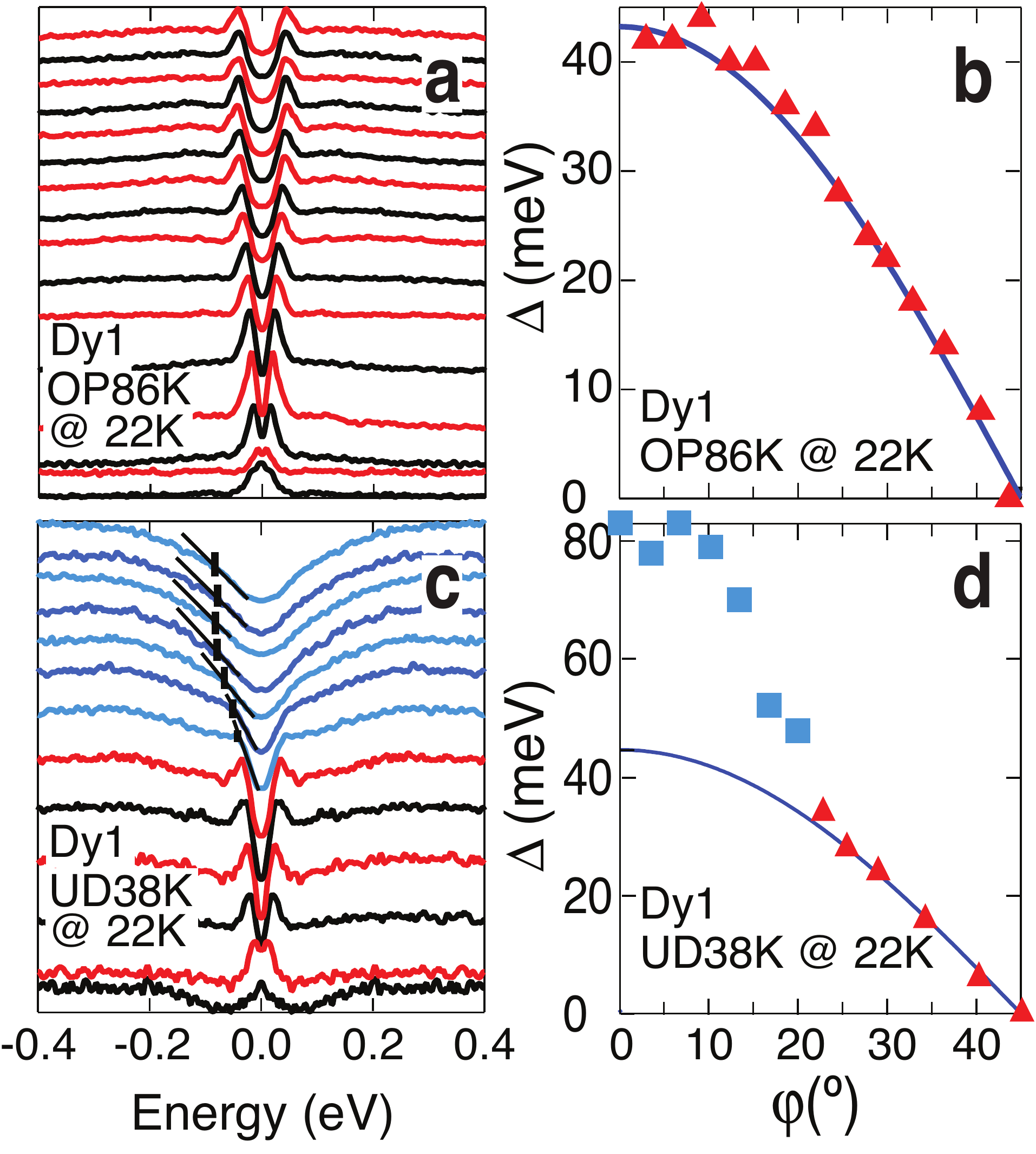}}
\caption{
Superconducting state spectra and energy gap  
for optimal and highly underdoped Dy1 samples. 
(a) Symmetrized spectra at ${\bf k}_F$, from the antinode (top) to the node (bottom) for 
an OP 86K Dy1 sample. (b) The gap
as a function of Fermi surface angle ($0^\circ$ is the antinode and $45^\circ$ the node).
The blue curve is a $d$-wave fit to the data.
(c) The same as panel (a) for an UD 38K Dy1 sample. Curves, near the node, 
with discernible quasiparticle peaks are shown in red, while those near the antinode are shown in
blue. (d) Gap along the Fermi surface from data of panel (c). See text for discussion of
the lines drawn in panel (c) and of the triangles (squares) used in panel (d) to denote whether there is (or is not) a quasiparticle peak in the spectrum.
}\label{fig2}
\end{figure}

\begin{figure}
\centerline{\includegraphics[width=.6\textwidth]{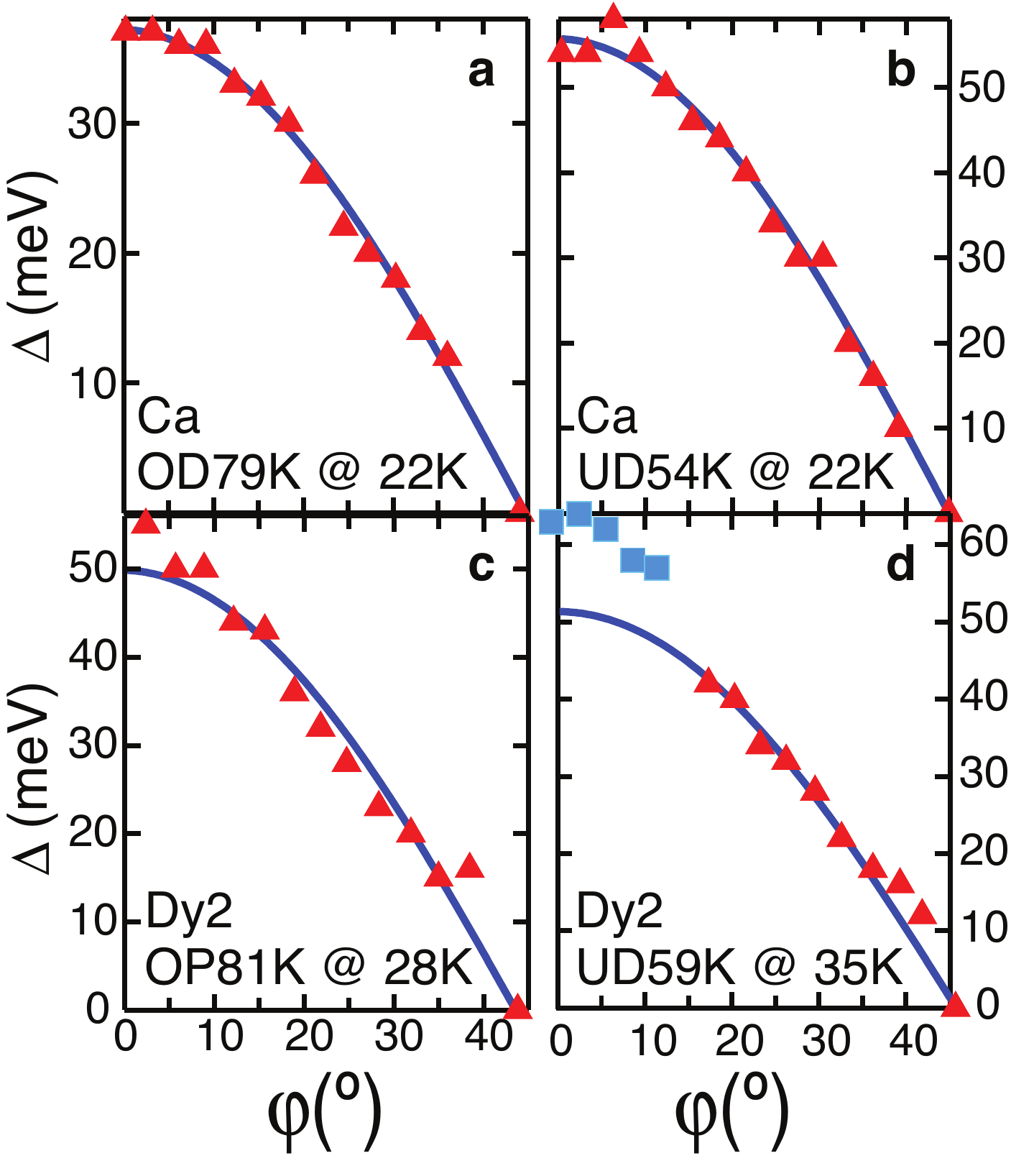}}
\caption{
Energy gap anisotropies of various samples. (a) OD 79K Ca (where $T_c^{\rm max} = 82$K);
(b) UD 54K Ca; (c) OP 81K Dy2; and (d) UD 59K Dy2. The two near optimal samples
in panels (a,c) both show a simple $d$-wave gap. This behavior persists in the UD Ca sample of panel
(b), but the UD Dy2 sample of panel (d) has a two-gap behavior (see text) despite
having a $T_c$ similar to the UD Ca sample. 
}\label{fig3}
\end{figure}

\begin{figure}
\centerline{\includegraphics[width=.6\textwidth]{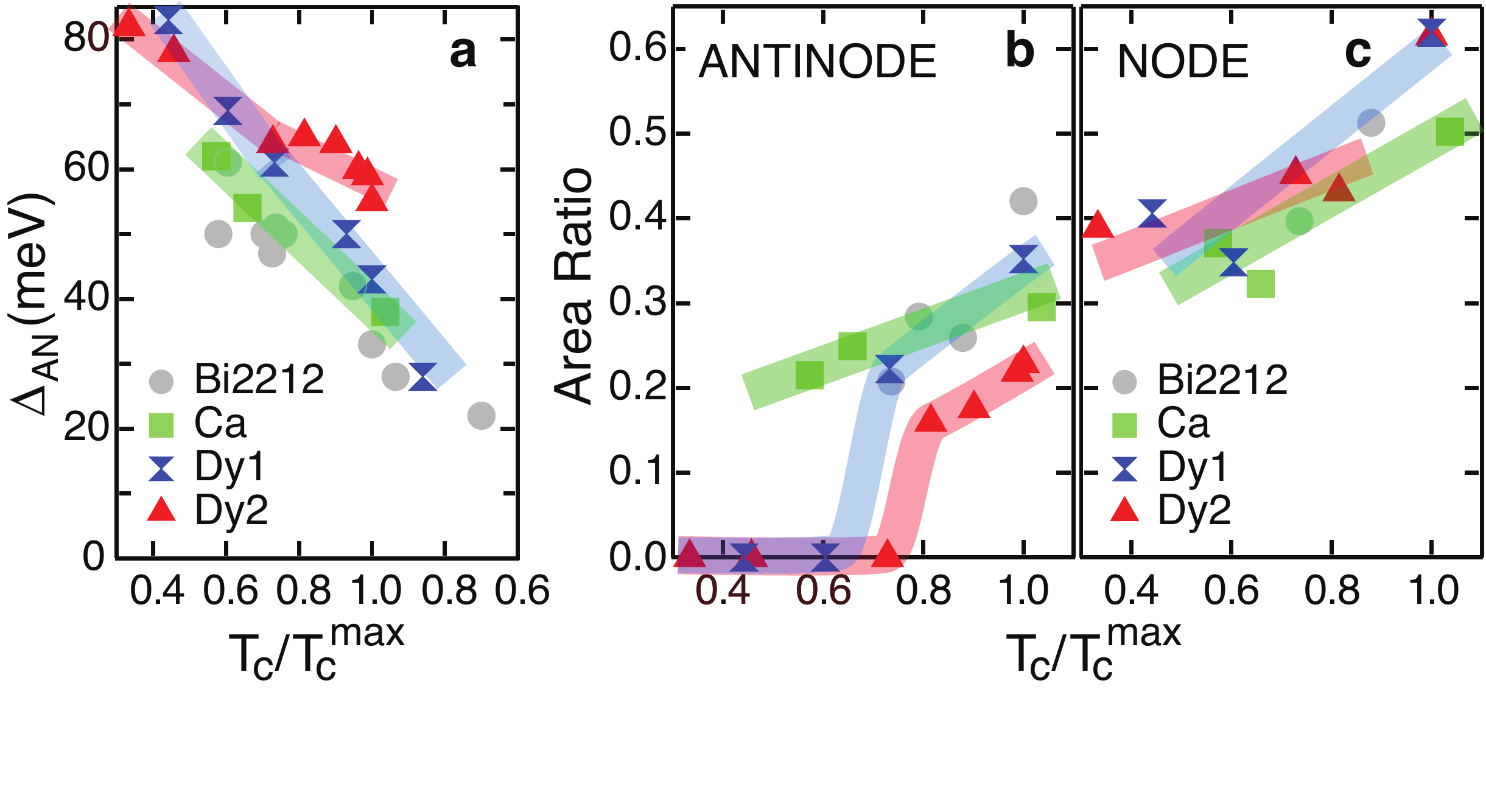}}
\caption{
Antinodal gaps and quasiparticle weights.
(a) The antinodal energy gap as a function of doping for various samples is seen to grow monotonically
with underdoping. In this and the next two panels, the doping is characterized by the measured quantity 
$t = T_c/T_c^{\rm max}$, with UD samples shown to the left of $t=1$ and
and OD samples to the right. All results are at temperatures well below $T_c$.
(b) The coherent spectral weight for antinodal quasiparticles as a function of doping. The Dy-doped 
samples exhibit a rapid suppression of  
this weight to zero for UD $T_c/T_c^{\rm max} < 0.7 - 0.8$, while the Ca-doped samples
show robust antinodal peaks even for $T_c/T_c^{\rm max} \simeq 0.5$.
(c) The coherent spectral weight for nodal quasiparticles 
as a function of doping, which is seen to be much more robust than the antinodal one.
}\label{fig4}
\end{figure}

\end{document}